# Effect of Rootstock on Some Aspects of Pistachio (*Pistacia vera* L.): A Review

Aram Akram Mohammed ✉ iD
*Horticulture Department, College of Agricultural Engineering Sciences, University of Sulaimani, Sulaimani, Kurdistan Region, Iraq*

Fakhraddin Mustafa Hama Salih iD
*Horticulture Department, College of Agricultural Engineering Sciences, University of Sulaimani, Sulaimani, Kurdistan Region, Iraq*

**Abstract:**

Budding and grafting are the strategies employed to combat unfavorable environmental conditions and improve some physiological defects in the *Pistacia vera* tree. Drought and salinity stresses are the most prominent adverse conditions encountered in pistachio production. It has been observed in different studies that various pistachio rootstocks can be used to ameliorate the effect of those two stresses. Besides, rootstock has a role in some physiological performances of pistachios such as nutrient uptake and photosynthesis. Furthermore, nut blank, unsplit nut, and alternate are three physiological disorders found in pistachio. Relationships have been found between the degree of these physiological disorders and the rootstock effect. The impact of rootstock on drought and salinity stresses, physiological performances, and physiological disorders in *P. vera* will be discussed in this review.

## Introduction

Pistachio (*Pistacia vera*) is essentially produced via budding and grafting onto a convenient rootstock, on the one hand, to obtain true-to-type seedlings with a determined sex type at the early stages of growth. This is because *P. vera* is a dioecious species, has a long vegetative stage, and there is no reliable marker to distinguish the sex type of the seedlings at the initial growth stages (Bai et al., 2019). Also, rooting of the *P. vera* cuttings is difficult and not vailed on a large scale (Noori et al., 2019). Thus, to generate true-to-type trees, budding and grafting are applied (Moriana et al., 2018). On the other hand, to combat unfavorable environmental conditions that may obstacle the production of *P. vera*, budding and grafting is used onto a resistible rootstock. However, *P. vera* is highly cultivated in arid and semi-arid areas, but the resistance of the different cultivars and genotypes to adverse conditions is variable (Hakimnejad et al., 2019). In contrast, there are certain cultivars of *P. vera* can withstand some of the tough conditions and can be used as a rootstock for the susceptible cultivars (Sheikhi et al., 2019). Besides, some other species in the *Pistacia* genus compatible with *P. vera* can be used as rootstock which have different degrees of resistance to adverse circumstances (Rahemi and Tavallali, 2007). Selection of a certain rootstock is a key component of *P. vera* cultivation. Various features of a tree's growth, productivity, and adaptability to different conditions are affected by the choosing of rootstock (Mir et al., 2023). This selection inevitably relies on the nature of the problem *P. vera* cultivation has, or the site and the country production. Several *P. vera* cultivars are employed as rootstock in Iran, such as 'Badami', 'Sarakhs', and 'Qazvini', and in Turkey *P. vera* is also used in most cases (Acar et al., 2017). The reasons that force the growers to use *P. vera* as a rootstock are the survival rate after transplanting is high due to lateral roots that are fully grown, and the seedlings of this species reach budding and grafting size in a shorter period. Whereas, in the USA, more specifically in California, *P. vera* has been completely replaced by other rootstocks in the last two decades due to susceptibility to *Verticillium*, thus *P. integerrima* and 'UCB-I' rootstocks have frequently been used for their tolerance to *Verticillium* (Kaska et al., 2002).

Growth and productivity of pistachio and other fruit trees are influenced by rootstock (Font et al., 2020). The relationship between scion and rootstock may reflect on variable responses originating from physiological and genetic factors (Warschefsky et al., 2016). To improve adaptation and productivity, the best tool is the diversification of cultivars and rootstocks. The main focus of recent studies is on the evaluation of pistachio rootstocks and scion–rootstock behavior from physiological and agronomic perspectives and their consequences on agro-food characteristics, under distinct environments (Noguera-Artiaga et al., 2020a). According to the interaction of scion-rootstock revealed in experiments, the responses were not consistent and occasionally contradicting (Ouni et al., 2022). This review aims to show the effect of rootstocks on some physiological aspects and adaptation for adverse conditions of *P. vera* that were obtained in numerous researches.

## Response of *P. vera* Rootstocks to Drought Stress

The growth and productivity of plants are negatively affected by drought stress (Anjum et al., 2011). Comparing with other environmental factors, drought stress resulted in the maximum obstruction (Shao et al., 2009). *P. vera* tolerates drought at a high level and can be cultivated under rainfed in several regions (Tadros et al., 2021). However, irrigation imparts better nut quality, a high shell splitting rate, and minimizes alternate bearing (Goldhamer and Beede, 2004). Managing good production of *P. vera* in drylands at all stages is an important issue (Fereres et al., 2003). Using efficient rootstock is one of the strategies to reduce the undesirable consequences of water deficit (Opazo et al., 2020). There are expectations that tolerable rootstocks contribute to the growth and yield of scion during water scarcity (Serra et al., 2014). Grafted *P. vera* onto *Pistacia atlantica, Pistacia integerrima,* and *Pistacia terebinthus* rootstocks were subjected to three irrigation treatments (regulated deficit irrigation (RDI) at stage two (shell hardening), i) to avoid water stress the grafted trees were regularly irrigated. ii) suppression irrigation until stem water potential (SWP) reached below -1.3 MPa, after that irrigated to maintain SWP below but near this threshold. iii) Irrigation to keep SWP at -1.5 MPa. It was observed that *P. atlantica* resulted in better yield and quality in comparison with the other rootstocks (Carbonell-Barrachina et al., 2015). Moreover, budded *P. vera* cv. Kerman onto *P. terebinthus*, *P. atlantica,* and *P. integerrima* were evaluated for their resistance to RDI in shallow soils at stage two by midday stem water potential ($\psi_{stem}$). The recommended stress was determined around -1.5 MPa, but at -2.0 MPa will be detrimental to pistachio production in the long term. *P. atlantica* followed by *P. terebinthus* were better than *P. integerrima*, and the worst physiological status was found with the latter one (Memmi et al., 2016). In a somewhat similar to previous results, Noguera-Artiaga et al. (2020b) exposed budded *P. vera* onto *P. terebinthus*, *P. atlantica* and *P. integerrima* rootstocks to RDI at SWP −1.5 MPa and <−2.0 MPa at stage two. In conclusion, they recommended that *P. terebinthus* at SWP −1.5 MPa was the best in terms of nut composition, functionality, and sensory quality. Apart from the RDI, Gijón et al. (2010) withdrew irrigation from 1-year-old 'Kerman' seedlings grafted on *P. terebinthus* L., *P. atlantica*, and a hybrid of *P. atlantica*×*P. vera* on

649

"day of the year" (DOY) 204 till 218 in pots. The rootstocks had an apparent role in the response of the Kerman cultivar to the water stress. A higher degree of stomatal control and low leaf senescence were observed in the seedlings grafted onto the hybrid. *P. terebinthus* encouraged powerful stomatal control and rapid leaf senescence. *P. atlantica* had similar levels of water stress as the two other rootstocks. In another pot experiment, 'Kerman' scions were budded onto 'UCB-I', *P. atlantica,* and *P. terebinthus* rootstocks and subjected to water stress for 28 days outdoor, in the second year after budding. The data showed that the three rootstocks gave rise to osmotic adjustment. Furthermore, the parameters confirmed that *P. atlantica* was more tolerant to drought, which could be due to greater activity of the roots. Whereas, the vegetative growth data implied 'UCB-I' was resistant more than *P. atlantica*. This may be because one of the parents of 'UCB-I' is *P. atlantica*, thus having an absorption capacity greater than *P. atlantica* (Moriana et al., 2018). Among the rootstocks belonging to *P. vera* species, 'Qazvini' had a better tolerance to drought than 'Badami' (Bagheri et al., 2012).

## Adaptation of *P. vera* to Salinity Through the Application of Rootstocks

Salinity is another issue that will face pistachio production worldwide. The majority of *P. vera* all over the world is cultivated in saline soils, or low quality and saline water is used for irrigation (Dehghani et al., 2023). Salinity causes deleterious effects on plants by inducing osmotic and specific-ion toxicity (Abbaspour et al., 2012). The worst outcomes that could occur in plants because of salinity are growth reduction, nutritional deficiencies or imbalances, toxicity symptoms, and reduction in the quality and quantity of fruit (Zörb et al., 2019). In the response of plants to soil salinity, rootstock has a crucial role because the first plant organ facing salinity is the root. Therefore, having precise knowledge about scion/rootstock interaction is a key strategy to avoid the injurious effects of salinity (Ahmad and Anjum, 2020). In many fruit tree species, sensitive scions to salinity are grafted onto the salt-tolerant rootstocks to combat salinity (Yin et al., 2010). In many different studies, the ability of *P. vera* rootstocks has been investigated to reveal the ability to withstand salinity stress. In this context, salinity stress of combined $SO_4^{2-}$, $Cl^-$, and boron (B) at 3.5, 8.7, 12, or 16 dS·m$^{-1}$ electrical conductivity (ECiw) was evaluated in *P. vera* cv. Kerman grafted onto *P. atlantica*, *P. integerrima,* and 'UCB-I' rootstocks. The salinity treatments were added through irrigation. The performance of 'Kerman' pistachio was significantly better onto *P. atlantica* and 'UCB-I' rootstocks than on *P. integerrima* at 16 dS·m$^{-1}$ (Ferguson et al., 2002). Besides, 'Qazvini', 'Sarakhs', and 'Badamin-e-zarand' rootstocks which belong to *P. vera* species were subjected to salinity stress after treatment with 0, 75, 150, or 225 mM NaCl in 8-L polyethylene pots via irrigation every 3 days. The data were calculated at 0, 30, and 60 day intervals. The most salt-tolerant rootstock was 'Qazvini' in comparison to 'Sarakhs' and 'Badami-e-zarand' rootstocks (Hokmabadi et al., 2005). In another similar study, Karimi et al. (2009) added 0, 800, 1600, and 3200 mg NaCl kg$^{-1}$ soil to the pots of 30-day-old 'Qazvini' and 'Badami' seedling rootstocks. The seedlings were irrigated with distilled water at 7-day interval after adding the NaCl amounts each time. The analysed data affirmed that 'Qazvini' was more resistant to salinity than 'Badami'. Panahi (2009) applied iso-osmotic agents (150 mM NaCl and 20% PEG-6000 solutions) by means of irrigation to *P. atlantica* subsp. *mutica*, *P. vera* 'Sarakhs', and *P. vera* 'Badami' rootstocks at 3-month-old. Iso-osmotic treatments caused 2-fold ABA in *P. atlantica* subsp. *mutica*, and *P. vera* 'Badami', but it was 1.5-fold in *P. vera* 'Sarakhs'. In addition, 'Badami-e-Zarand A', 'Badami-e-Zarand B', 'Qazvini', and 'Sarakhs' rootstocks were irrigated with saline water at 0.75, 5, 10, and 15 ds.m$^{-1}$. The assessment of salinity tolerance established that 'Badami-e-Zarand B' could be used as a salt-tolerant rootstock, and 'Sarakhs' was a sensitive rootstock (Karimi et al., 2011). Among the subspecies of *P. atlantica*, *P. atlantica (standard)* as resistant, and *P. atlantica* subsp. *Kurdica* as sensitive to salinity was determined under 0.75, 5, 10, and 15 dS.m$^{-1}$ (Karimi, et al., 2012). In their study, Karimi et al. (2014) selected the hybrid rootstock ('Badami-Riz-e-Zarand'-Female×*P.atlantica*-male) as the best rootstock for salinity over 'Qazvini' and 'Badami-Riz-e-Zarand' rootstocks under 0, 60, and 120 mM NaCl+CaCl$_2$+MgCl$_2$ (3:2:1) for 45 days. Furthermore, *P. vera* and *P. atlantica* were employed to adapt Mateur cultivar to salinity. The grafted trees were irrigated with three levels of saline water (tap water (EC$_w$:1.95 dSm$^{-1}$), moderately saline water (EC$_w$: 5 dSm$^{-1}$), and highly saline water (EC$_w$: 12 dSm$^{-1}$), three times a month for three years. The findings explained that *P. atlantica* rootstock was an outstanding option to avoid *P. vera* cultivars from saline stress. This may be because *P. atlantica* capacity for nutrient uptake, control, and accumulation of salt ions (Na$^+$ and Cl$^-$), selectivity for K$^+$-Na$^+$ and Ca$^{2+}$-Na$^+$, and resuming normal growth and fruiting after relieving from saline stress are high (Mehdi-Tounsi et al., 2017). Six rootstocks for *P. vera* ('Badami', 'Akbari', 'Italyayi', 'Ahmad-Aghaee', 'Qazvini', and 'UCB-I') were tested for their resistance to salinity stress. The salinity treatment was NaCl at 0.5, 12, and 18 dSm$^{-1}$. 'UCB-I' and 'Akbari' followed by 'Ahmad-Aghaee' had better performance in the saline conditions (Raoufi et al., 2019). Also, Goharriz et al. (2020) recognized 'UCB-I' and 'Badami' as the most salinity- and drought-resistant rootstocks (35% of field capacity + 250 mM NaCl) condition, 'Qazvini' as moderate, and 'Kalleh-ghuchi' as sensitive. Additionally, the type of salt determines the severity of the stress. Adish et al. (2010) studied the effect of 0, 100, and 200 mM NaCl and CaCl$_2$ on pistachio rootstock 'Badami' at the seedling stage. A negative salinity effect was observed owing to both NaCl and CaCl$_2$, however impact of NaCl was more extreme than CaCl$_2$. In spite of these studies, the salinity tolerance of rootstocks for *P. vera* has been examined *in vitro*. Chelli-Chaabouni et al. (2010) culture seeds of *P. vera* and *P. atlantica* *in vitro* in media supplemented with low NaCl at 0 to 80mM for 45 days, or high NaCl at 0, 131, and 158.5mM for *P. vera* and 0, 131, and 240mM for *P.atlantica*. The lethal concentrations were 158.5 mM for *P. vera* and 240 mM for *P.atlantica*.

*P.atlantica* controlled sodium and chloride uptake more efficiently than *P. vera*. In an additional *in vitro* investigation, the multiplied shoots of 'Akbari', 'Ahmad-Aghaee', 'Italyayi', 'Badami', 'Qazvini', and 'UCB-I' pistachios were grown in media containing 0, 60, and 120 mM NaCl. 'Akbari' had the best performance at 120 mM NaCl (Raoufi et al., 2021).

## Nutrient Uptake is Affected by Rootstock in *P. vera*

Plant nutrition is regulated by soil, environment and plant factors. Among them, plant factors associated with variety, age, growth stage, and root structure (Erdal and Nazli, 2019). The researches have been explained that rootstock is another plant factor that plays a role in the uptake of nutrients (Valverdi and Kalcsits, 2021). Mineral nutrition of the grafted plant can be enhanced due to rootstock. For that reason, employing efficient rootstock may ameliorate mild nutrient deficiency and minimize the application of nutrients in some cases (Ibacache and Sierra, 2009). The level of nutrients in scion is variable depending on the type of rootstock, and this variability likely originates from inter-specific variation among rootstocks with regard to nutrient absorption and translocation to scion (Vahdati et al., 2021). Caruso et al. (2003) demonstrated that *P. atlantica* rootstock delivered more N, P, K, Ca, Mg, Fe, Mn, Zn, and Cu to *P. vera* cv. Bianca scion than *P. integerrima* and *P. terebinthus* rootstocks. In *P. vera*, the type of rootstock governs the nutrient content and the type of the mineral in leaf and kernel of the nuts. Tavallali and Rahemi (2007) analyzed the nutrient amounts in the leaf and kernel of 'Ohadi', 'Kalleh-ghuchi', and Ahamd-aghaii grafted onto *P. vera* cv. Badami and Sarakhs, and *P. atlantica* subsp. *mutica* rootstocks. The highest P, Zn, K, and lower Na and Mg were measured in the leaves of the cultivars grafted onto Beneh rootstock. 'Badami' resulted in higher leaf Ca and lower Zn. Higher leaf Fe and Cu were detected in the grafted trees onto 'Sarakhs'. Besides, kernel nutrient contents (P, K, Mg, Fe, Zn, and Cu) of the *P. vera* cultivars were higher onto 'Sarakhs' rootstock. Sherafati et al. (2009)

found that the acquisition of potassium and zinc by 'Akbari' scion was the best onto 'Badami' rootstock, but the worst onto Daneshmandi rootstock. The highest amount of iron into 'Barg-seyah' scion was onto 'Kalleh-ghuchi' rootstock. Moreover, the *P. vera* rootstocks are different in response to symbiotic fungi to improve nutrient uptake. Bagheri et al. (2012) observed that 'Qazvini' rootstock was better at absorbing P and Mn than 'Badami' under *Glomus mosseae* inoculation. Meanwhile, the same rootstock does not accumulate the same nutrient in different scions at similar concentration. At the beginning of fruit maturation, Surucu et al. (2020) collected leaf samples from many *P. vera* cultivars grafted onto *Pistacia khinjuk* rootstock and analyzed for Mg, Ca, N, P, K, Fe, Cu, Mn, and Zn contents. The highest N, P, and K in Haciserifi, Mg, Ca, and Cu in 'Mumtaz', Fe and Zn in 'Vahidi', and Mn in 'Sel-15' were recorded.

## Relationship Between Rootstock and Photosynthesis Performance in *P. vera*

The photosynthetic rate of scion varieties is distinctly influenced by rootstock. Rootstock can control canopy size, leaf morphology, and gas exchange which in turn affect photosynthesis competence (Losciale et al., 2008). Thus, as a crucial criterion for selection, Jover et al. (2012) reported that the effect of rootstock on photosynthesis should be taken into account. Also, Fullana-Pericàs et al. (2020) referred to grafting onto a convenient rootstock can be a beneficial technique to make better photosynthetic performance in plants. In the case of *P. vera* rootstocks, the photosynthesis rate was assessed in 'Ohadi', 'Kalleh-ghuchi', and 'Ahmad-Aghaee' scions grafted on *P. vera* cv. Sarakhs and Badami riz, *P. atlantica* subsp. *atlantica* and *P. atlantica* subsp. *mutica* rootstocks. The trees grown onto *P. vera* cv. Sarakhs and *P. atlantica* subsp. *atlantica* had the maximum photosynthetic rates, in contrast, the rate of photosynthesis was the least in the trees their rootstocks were *P. atlantica* subsp. *mutica* and *P. vera* cv. Badami (Fotouhi et al., 2006). Scion-rootstock combinations of *P. vera* cv. Mateur and Kerman scions onto *P. vera* and *P. atlantica* rootstocks were evaluated by Ghrab et al. (2014). It was obtained that *P. atlantica* rootstocks provide the maximum photosynthesis and chlorophyll content for the two scions.

## Rootstock to Increase Nut Splitting Rate

One of the desirable characteristics of pistachio nuts is the splitting of the shell (longitudinal dehiscence). It is a much more attractive trait that encourages a variety choosing in the markets (Ouni et al., 2022). During the harvesting weeks, the amount of split nuts increased because of the enlargement of the inner kernel which forces the outer shell to open. However, after harvesting, a large quantity of closed nuts remains due to irrigation scheduling, imbalanced nutrition, fertilizer application, cultivar, salinity, and temperature upon blooming stage (Sukunza et al., 2023). Hence, special treatment should be conducted artificially on unsplit nuts aiming splitting them, such as longitudinal force on the shell suture and thermal processing (Foroutanaliabad and Foroutanaliabad, 2006). Nevertheless, these methods are expensive, require special equipment, and nut quality may be reduced. Regarding nut splitting factors, the rootstock is among the factors that have been reported to impact pistachio nut splitting (Ouni et al., 2018). Early nut splitting was examined by Tajabadipour et al. (2006) in 'Ohadi', 'Kalleh-ghuchi', and 'Ahmad-Aghaee' tree cultivars grafted onto *P. vera* cv. Badami-e-Riz, *P. vera* var 'Sarakhs', *P. atlantica* subsp. *mutica,* and *P. atlantica* subsp. *atlantica* rootstocks. 'Kalleh-ghuchi' onto *P. atlantica* subsp. *atlantica* gave the highest percentage of earlier splitting, but the same cultivar delayed nut splitting onto *P. vera* cv Badami-e-Riz rootstock. The rootstocks did not induce earlier nut splitting in 'Ohadi' and 'Ahmad-Aghaee'. In addition, three *P. vera* cultivars ('Ohadi', 'Kalleh-ghuchi', and 'Ahmad-Aghaee') were compared to their differences in nut quality including splitting rate onto *P. vera* cv. Sarakhs and Badami, and *P. atlantica* subsp. *mutica* rootstocks. The highest splitting percentage was recorded in the *P. vera* cultivars whose rootstock was *P. vera* cv. Badami (Rahemi and Tavallali, 2007). Effects of pistachio rootstocks (*P. vera*, *P. atlantica,* and *P. khinjuk*) on nut splitting of 'Siirt' and 'Ohadi' cultivars were examined. Shell splitting rate was outstanding in both cultivars as produced nuts onto *P. atlantica* rootstock, but the worst onto *P. khinjuk* rootstock. At the same time, the 'Siirt' cultivar was better than 'Ohadi' regarding nut splitting (Turker and Ak, 2010). Whereas, there is reports that nut splitting is characteristic of cultivar rather than the effect of rootstock. Parfitt et al. (2005) stated that cultivar and shell splitting are closely linked. Some *P. vera* cultivars were studied to determine the amount of nut splitting that grew onto *P. khinjuk* rootstock. It was obtained that Sel-2, Sel-5, and 'Siirt' cultivars had the highest percentage of split nuts related to 'Haciserifi', 'Kerman', Mumtaz, 'Ohadi', 'Sel-1', 'Sel-10', 'Sel-11', 'Sel-14', 'Sel-15', 'Uzun', and 'Vahidi' cultivars (Surucu et al., 2020). Additionally, Caruso et al. (2003) declared that rootstock did not significantly affect nut splitting, when they studied *P. vera* cv. Bianca onto *P. atlantica*, *P. integerrima*, *P. terebinthus* rootstocks.

## Role of Rootstock in Blank Nut in *P. vera*

A pistachio nut without a kernel is called a blank (empty) nut. This occurs as a result of failure of the embryo to develop, development of ovary tissues without successful fertilization, embryo abortion, tree inability to provide assimilates to complete kernel growth to a sufficient extent, and inadequate irrigation and boron leaf levels (Kashaninejad and Tabil, 2011). The severity of the blank is associated with cultivar, rootstock, and year to year (Hormaza and Wünsch, 2007). There are reports that rootstock lessens blank nut in *P. vera*. In this respect, *P. vera* cv. Ohadi, 'Kalleh-ghuchi', and 'Ahmad-Aghaee Ahmad-Aghaee' trees were investigated for blankness rate onto *P. vera* cv. Sarakhs and Badami, and *P. atlantica* subsp. *mutica*. *P. vera* cv. Badami rootstock produced the minimum of nuts with blankness (Rahemi and Tavallali, 2007). In another experiment, Turker and Ak (2010) showed that 'Siirt' and 'Ohadi' pistachio cultivars onto *P. vera*, *P. atlantica*, and *P. khinjuk* had various nut filling percentage. The lowest blank nuts were calculated from the two cultivars onto *P. atlantica* rootstock, on the contrary the two cultivars onto *P. khinjuk* exhibited the peak blank nuts. Whereas, not all studies found that rootstock determines the amount of blank nuts in pistachio. Caruso et al. (2003) announced the ineffectiveness of rootstocks (*P. atlantica*, *P. integerrima*, *P. terebinthus*) to change the ratio of blank nut significantly in *P. vera* cv. Bianca.

## Rootstock and Alternate Bearing in *P. vera*

The prominent obstacle in the production of pistachio is alternate bearing. It is a physiological disorder observed in pistachio and other fruit species in which a high yield quantity is obtained in one year (on year), but in the following year (off year) little or no yield may be harvested. Despite nut quantity, this disorder relates to nut quality from blank and non-split nut perspectives. In 'on' year, non-split nuts are profound and blanking is much more common in 'off' years (Kashaninejad and Tabil, 2011). Alternate bearing has been attributed to certain reasons. Competition between the developed inflorescence buds and heavy crop in the summer of 'on' year on carbohydrate may be one of the causes in which the produced nuts deprives the inflorescence buds from reaching sufficient carbohydrate and eventually abscised (Baninasab and Rahemi, 2006). On the other hand, the nutrient status of the tree is an additional factor that may influence the degree of alternate bearing. Kumar et al. (2016) found that the application of NPK and B decreased the flowering and fruit set along with reduction in bud abscission during 'on' and 'off' years. Warm climate and rain-fed conditions accentuate this phenomenon; also it is variable according to cultivar (Ghrab et al., 2014; Carbonell-Barrachina et al., 2015). The rootstocks may impact the intensity of alternating bearing in pistachio (Kaska et al., 2002). Pistachio rootstocks have different ability to uptake nutrient from the soil (Surucu et al., 2020), and as mentioned above nutrient is one of the factors that influences alternate bearing. Also, it was hypothesized that the vigor of pistachio rootstock mitigated alternate bearing because vigorous rootstock with large leaf surface area can manufacture more carbohydrates (Khezri et al., 2020). Beede et al. (2017) showed that Kerman *P. vera* cultivar onto 'PGII' and 'UCB-I' rootstocks had lower alternate bearing than onto *P. atlantica* and 'PGI' rootstocks. Lower bud abscission was detected in 'Kalleh-ghuchi' pistachio cultivars onto *P. vera* cv. Badami-Riz rootstock in comparison to *P. atlantica*, *P. atlantica* subsp. *mutica*, and *P. khinjuk* (Khezri, 2010). *P. vera* and *P. atlantica* rootstocks were tested for their effect on the functionality of *P. vera* cv. Mateur and Achouri in 2014 and 2015. Yield quantity indicated 2014 as 'on' year and 2015 as 'Off' year. In 2015, *P. atlantica* rootstock provided Mateur cultivar with a better yield than Achouri onto *P. vera* rootstock. In contrast, the impact of the rootstocks was insignificant in 2014 on both cultivars (Ouni et al., 2018). However, Ouni et al. (2022) referred that alternate bearing in *P. vera* cv. Mateur and Achouri onto *P. vera* and *P. atlantica* rootstocks varied depending on the year to year and cultivar but not rootstock, 'Mateur' was better than 'Achouri'.

## Conclusion

*P. vera* is a species that can tolerate different adverse conditions, but the degree of tolerance is different according to genotypes and cultivars. Rootstock has been observed that effective for adapting this species to unfavorable conditions. Through this review, it was revealed that *P. vera* 'Badami', 'Qazvini', 'Sarakhs', 'Akbari', and some other cultivars along with other species from the *Pistacia* genus such as *P. atlantica*, *P. terebinthus*, *P. integerrima*, *P. khinjuk,* and a hybrid of *P. atlantica*× *P. integerrima* ('UCB-I') can be used for different purposes. The studies on drought stress recommended *P. atlantica* and 'Qazvini' as drought-resistant rootstock. Also, *P. atlantica* and 'UCB-I' along with 'Qazvini' followed by 'Akbari' had the best performance in saline conditions. Regarding nutrient uptake, each rootstock had the best absorption capacity for specific nutrient elements, but, generally, *P. atlantica*, 'Badami', 'Sarakhs' rootstocks absorb the majority of the nutrient elements at a high level. The investigation indicated that rootstocks affected some physiological characteristics and physiological disorders of *P. vera*. In this context, *P. atlantica* and 'Sarakhs' caused the best photosynthesis. Besides. *P. atlantica* and 'Badami' resulted in the lowest blank and alternate bearing but the maximum nut splitting rate. However, some studies do not support the influence of rootstocks on nut splitting, blank nut, and alternate bearing in pistachio. Thus, rigorous investigations are needed to confirm the role of rootstocks in these physiological disorders.

## Conflict of Interests

No conflict of interest.